\documentclass[a4paper, 11pt]{article}
\date{December 2005}
\usepackage{epsfig}
\usepackage{latexsym}
\usepackage{amssymb}
\newcommand{\be}{\begin{equation}}
\newcommand{\ee}{\end{equation}}
\newcommand{\ba}{\begin{eqnarray}}
\newcommand{\ea}{\end{eqnarray}}
\newcommand{\bi}{\begin{itemize}}
\newcommand{\ei}{\end{itemize}}
\newcommand{\tr}{{\rm Tr\,}}
\newcommand{\re}{\mathop{\rm Re}}

\newcommand{\nn}{\nonumber \\}

\newcommand{\half}{{\textstyle\frac{1}{2}}}

\newcommand{\<}{\langle}
\renewcommand{\>}{\rangle}
\newcommand{\eq}{Eq.~}

\newcommand{\la}{\label}

\newcommand{\ello}{\ell_{0}}
\newcommand{\txts}{\textstyle}
\newcommand{\scrs}{\scriptstyle}
\newcommand{\hello}{\breve\ell_{0}}
\newcommand{\hro}{\breve r_{0}}
\newcommand{\hpsi}{\hat\psi}

\newcommand{\hchi}{\hat\chi}

\newcommand{\hM}{\breve M}
\newcommand{\hE}{\breve E}
\newcommand{\hmu}{\hat\mu}
\newcommand{\dBdloga}{\frac{d\beta}{d\log a}}
\newcommand{\dBdlogas}{\textstyle{\frac{d\beta}{d\log a}}}
\newcommand{\dkdloga}{\frac{d\kappa}{d\log a}}
\newcommand{\dkdb}{\frac{d\kappa}{d\beta}}
\newcommand{\bx}{\mathop{\bf x}}
\newcommand{\by}{\mathop{\bf y}}
\newcommand{\Tb}{\mathop{{\rm \bf T}_\beta}}
\newcommand{\dd}{{\rm d}}
\newcommand{\Kb}{\mathop{K_\beta}}
\newcommand{\db}{\mathop{\partial_\beta}}
\newcommand{\dk}{\mathop{\partial_\kappa}}

\begin{document}
\begin{titlepage}
DESY 06-145\\

\begin{centering}
\vfill

 \vspace*{2.0cm}
{\bf \Large Sum Rules and Cutoff Effects in Wilson Lattice QCD}

\vspace{3.0cm}
{\bf Harvey~B.~Meyer}
\centerline{Deutsches Elektronen-Synchrotron DESY}
\centerline{Platanenallee 6}
\centerline{D-15738 Zeuthen}
\vspace{0.1cm}\\
\centerline{harvey.meyer@desy.de}

\vspace*{4.5cm}

\end{centering}
\centerline{\bf Abstract}
\vspace{0.1cm}
\noindent 
We use the transfer matrix formalism to derive non-perturbative
sum rules in Wilson's lattice QCD with $N_{\rm f}$ flavours of quarks. 
The discretization errors on these identities are treated in detail.
As an application, it is shown how the sum rules can be exploited
to give improved estimates of the continuum spectrum and static 
potential.

\vspace{2.0cm}
\vfill
\end{titlepage}

\setcounter{footnote}{0}
\section{Introduction\la{sec:intro}}
Although relatively old, the subject of lattice sum rules~\cite{michael2,rothe1}
in $d=4$ SU($N$) gauge theories
has received only sporadic attention since it was initiated by 
C.~Michael~\cite{michael1}. 
In essence, these identities relate the derivative with respect to a bare
lattice parameter of a physical quantity to a higher order correlation function.
A prototype sum rule states that the $\beta$-dependence of an energy level 
is determined by the expectation value of the Lagrangian on that state.

In Monte-Carlo simulations, sum rules were first applied 
as a cross-check in form factor calculations 
at zero-momentum transfer~\cite{michael-tickle}, and in measurements 
of the flux-tube profile in the static $Q\bar Q$ system~\cite{michael-pot}. 
Later they were used to determine lattice 
beta-functions non-perturbatively~\cite{michael-beta,schilling-beta}.
On the theory side, they have also led to some insight into the relative 
contribution to the mass of a glueball
of the component $T_{00}$  of the traceless energy-momentum tensor
and of the trace anomaly~\cite{rothe2}.

Here we shall generalize the sum rules to the case where $N_{\rm f}$ flavours
of Wilson quarks are present. We do so by invoking the transfer matrix,
in our view the most elegant approach. It also makes it clear
that the identities derived do not depend on boundary conditions or 
on which phase the system is in --- 
be it a finite-temperature phase, a parity-broken phase or other.
Let us recall the Wilson action, which has two parameters $(\beta,\kappa)$:
\ba
S_{\rm g} &=& \beta \sum_p \square_p,\qquad  
\square_p=\textstyle{\frac{1}{N}}\re\tr\{1-U(p)\},\la{eq:Sg} \\
S_{\rm f} &=& \sum_x \bar\psi{\scrs(x)} \psi{\scrs(x)}
- \kappa \textstyle{\sum_\mu } \Big[
\bar\psi{\scrs(x)}U_\mu{\scrs(x)}(1-\gamma_\mu) \psi{\scrs( x+a\hmu)} + 
\bar\psi{\scrs( x+a\hmu)}U_\mu^\dagger{\scrs (x)}(1+\gamma_\mu) \psi{\scrs(x)} \Big]
\la{eq:Sf}
\ea
where in \eq\ref{eq:Sg} the sum extends over all unoriented plaquettes.

It is well-known that cutoff effects are of O($a$) in Wilson's formulation.
In recent times, most  large-scale numerical calculations have made use 
of the O($a$) improvement program to reduce these effects
(with some notable exceptions~\cite{luscher-LAT05,alexandrou}). 
We present an idea to exploit lattice sum rules in order to extract
additional information on the lattice spacing dependence of the spectrum, and further 
how its lattice artefacts  can be reduced.

The idea is also applicable to the energy levels in the presence of a $Q\bar Q$ pair,
where one is typically interested in their derivatives with respect to the 
static quark separation. The static force, and the effective central charge
in the pure gauge case, are affected by discretization errors 
which can be large compared to the statistical precision one is able to achieve.

In section 2 we rederive some of the known results in the pure gauge theory, 
where  particular attention is paid to discretization errors. 
In section 3  the sum rules are derived for full QCD.
Concrete applications are proposed in both sections. We make some 
concluding remarks in section 4.
\section{Transfer matrix and sum rules in the pure gauge theory}
We start with a brief reminder on the transfer formalism and how sum rules
are derived in the pure gauge theory. Some familiarity with the lattice
regularization is assumed. For dimensionful quantities in lattice units
we use the notation $L=\breve L a$ etc.

If $V$ is the set of spatial link variables
in a time-slice, let $\Phi[V]\in{\cal H}_G$ be a square-integrable wave function with 
respect to the Haar measure of the gauge group.
The transfer matrix $\Tb$ acts as follows~\cite{luscher-les-houches,luscher-transfmat,creutz}:
\ba
(\Tb \Phi)[V]&=&\int \prod_{\bx} \txts{\prod_{k=1}^3}~ \dd V_k'(\bx)~ \Kb[V,V']~  \Phi[V'],
\la{eq:ker} \\
\textrm{with  kernel}\quad
\Kb[V,V']&=&\int \prod_{\bx} \dd W(\bx)~ \exp\left(-\beta\Delta S[V,W,V']\right),\\
\Delta S[V,W,V'] &=& \sum_{\bx} \Big\{
\txts{\sum_k}  \square_{0k}[V,W,V'] + \frac{1}{2} \txts{\sum_{k<l}}
 (  \square_{kl}[V] +  \square_{kl}[V'] )  \Big\}\\
  \square_{0k}[V,W,V'] &\equiv& \frac{1}{N}\re\tr
      \{1-V_k'(\bx)W(\bx+a k)V_k(\bx)^{-1} W(\bx)^{-1}\}.
\ea
In words, $\Delta S$ is the restriction of the action to the region of space-time
between $x_0$ and $x_0+a$, with a weight of 1/2 given to the terms living on the boundary.
When the full gauge system is defined with periodic boundary conditions in the time
direction, the partition function is given by $ {\cal Z}(\beta) = \tr \{ {\Tb}^{\breve L_0}\}$.

We define the magnetic-plaquette operator $\hat \square_{kl}$
to act as in Eq.~\ref{eq:ker} with  $\Kb$ replaced by
\be
K_{kl}[V,V']=\delta(V-V')~  \square_{kl}(0,\bx)[V].
\ee
The kernel corresponding to the electric plaquette operator $\hat\square_{0k}$ is defined as
\be
K_{0k}[V,V']= \int \prod_{\bx} \dd W(\bx)~ \exp\left(-\beta\Delta S[V,W,V']\right)~
 \square_{0k}[V,W,V'].
\ee
Then, if ${\bf T} |\Phi\> = \lambda |\Phi\>$,
\be
-\db \lambda(\beta) = -\db \<\Phi |  {\bf T}_\beta |\Phi\>=-\<\Phi | \db  {\bf T}_\beta |\Phi\>
= \lambda(\beta) \<\Phi|\txts{ \sum_{k<l}}\hat\square_{kl} | \Phi\> +
                        \<\Phi| \txts{\sum_k} \hat\square_{0k} | \Phi\>.
\ee
Since $|\Phi\>$ can be any eigenstate of the transfer matrix, we can choose
it to be successively the vacuum and a glueball state.
The spatial volume is assumed to be large enough for 
the finite-volume effects on the mass gap to be negligible.
Subtracting the two corresponding rules gives
\be
\frac{d\hM}{d\beta}  = \txts{\sum_{k<l} }\<\Phi|\hat \square_{kl}\Phi\>-\<\Omega|\hat \square_{kl}\Omega\> 
~+~ \txts{\sum_k} e^{\hM}\<\Phi| \hat \square_{0k}|\Phi \> -e^{\hE_\Omega} \<\Omega| \hat \square_{0k}|\Omega \>.
\la{eq:result}
\ee
Here $\hM=-\log\lambda/\lambda_\Omega$ is the mass of the glueball in lattice units.
One might be surprised to see the vacuum energy appear in this expression, but 
as we shall see shortly, computing the expectation value of an electric
plaquette in the usual ensemble yields precisely $e^{\hE_\Omega} \<\Omega| \hat P_{0k}|\Omega \>$.

Following~\cite{michael2}, we can use the fact that $d/da(\hM/a)=0$
in the continuum limit to conclude
\be
\hM = \dBdloga \Big\{\txts{ \sum_{k<l}} \<\Phi|\hat \square_{kl}|\Phi\> 
- \<\Omega|\hat \square_{kl}|\Omega\>
~+~  \txts{ \sum_k} e^{\hM}\<\Phi| \hat \square_{0k}|\Phi\> 
 - e^{\hE_\Omega}\<\Omega| \hat \square_{0k}|\Omega \>\Big\}
[1+{\rm O}(a^2)].
\ee

To translate the matrix elements appearing on the right-hand side
into Euclidean correlators, for $\Phi$ the lightest state in its
symmetry channel one may choose a zero-momentum
linear combination $\phi(x_0)$ of magnetic loops in that channel, 
and compute the correlator
\be
\frac{\< \bar \phi(x_0+a)~ 
\half\Big(\sum_{k<l} P_{kl}(x_0/2)+ P_{kl}(x_0/2+a)\Big)  ~\bar \phi(0) \> }
{ \< \bar \phi(x_0+a)~ \bar \phi(0) \> } ~\stackrel{x_0\to \infty}{\to}~
\sum_{k<l} \<\Phi|\hat \square_{kl}|\Phi\>,
\ee
where $\bar \phi(x_0) = \phi(x_0) - \< \phi\>$ (take $x_0/a$ even).
And similarly
\be
\frac{\< \bar \phi(x_0+1)~ \sum_{k} P_{0k}(x_0/2)  ~\bar \phi(0) \> }
{ \< \bar \phi(x_0+a)~ \bar \phi(0) \> } ~\stackrel{x_0\to \infty}{\to}~
 e^{\hM(\beta)} \sum_k \<\Phi| \hat \square_{0k}|\Phi \>.
\ee
\subsection{Improved estimators of the continuum spectrum\la{sec:impr}}
We use the shorthand notation $\frac{d\hM}{d\beta} = \square^\Phi_\Omega$ 
for ~\eq\ref{eq:result} from now on.
In view of taking the continuum limit, we use a reference 
length $\ello$ which sets the scale of the theory. Typical examples
are $\sigma^{-1/2}$, where $\sigma$ is the string tension and $r_0$, 
the Sommer reference scale~\cite{sommer}.
We assume that for every $\beta$, $\hello\equiv\ello/a$ 
can be determined, so that $\beta(\hello)$, which is assumed to be 
monotonic in the range of interest, and 
\be
\dBdloga \equiv -\frac{d\beta(\hello)}{d\log \hello}
\la{eq:dBdloga}
\ee
are well-defined. For the case $\ello=r_0$, $\hello(\beta)$ is known
with one percent precision or better~\cite{necco-sommer}.
Thus on a line of constant physics, a function
of $\beta$ can just as well be regarded as a function of $\hello$, 
and vice versa.

The dimensionless quantity $z  \equiv     \ello M$
has a continuum limit, and we can express its derivative 
with respect to the lattice spacing exactly:
\be
\frac{dz}{d\hello} = \hM - \dBdloga~ \square^\Phi_\Omega.
\la{eq:dzdhello}
\ee
This information on the slope can be included in the continuum extrapolation
if data at several lattice spacings is available. Even if not,
it can  be used to provide an improved estimate of the continuum limit
of $z$ if we assume a particular form of the discretization errors.
In practice $z$ is usually observed to 
approach the continuum with O($a^2$) corrections within statistical 
errors; such corrections can be removed according to:
\be
z_{\rm impr} \equiv z(\hello) + \half\hello\frac{dz}{d\hello} 
                  = \frac{3}{2} z(\hello) - 
                   \frac{\hello}{2}\dBdloga~  \square^\Phi_\Omega.
\la{eq:zimp}
\ee
In what sense is this an improved estimate of the continuum limit?
Suppose the true analytic form of $z$ is~\cite{symanzik,advanced-qcd}
\be
z(\hello) = z_{\rm cont} + \frac{1}{\hello^2}
                           \sum_{n=0}^N c_n \log^n\{1/\hello\}
                         +{\rm O}(1/\hello^4).
\ee
Then
\be
z_{\rm impr} = z_{\rm cont} - \frac{1}{2\hello^2} \sum_{n=0}^{N-1}
                              (n+1) c_{n+1} \log^n\{1/\hello\}
                            + {\rm O}(1/\hello^4).
\ee
Since the series of logarithms is asymptotic, if $N=1$ yields
the best accuracy at the lattice spacing one is working at, then
$|c_0| \gg |c_1 \log\{1/\hello\}|\gg |c_1|$, so that $z_{\rm impr}$
has reduced discretization errors, although it is not a full O($a^2$) improvement.
Note that it is essential in (\ref{eq:zimp}) 
to use the quantity $\ello$ that appears elsewhere in the formula 
to define $\dBdloga$, as in~\eq(\ref{eq:dBdloga}).

\subsection{States depending on an external length scale}
We now consider an eigenstate of a transfer matrix which, in 
addition of depending on $\beta$, also depends on a physical length scale 
$\ell$, and the latter can take only integer multiples of some quantum $\Delta s$:
\be
s = n~ \Delta s, \qquad n\in {\bf N}.
\ee
Examples are zero-momentum  states in finite volume
(where $s=L,~\Delta s=a$), but also large-volume states with non-vanishing momentum
($s=p,~\Delta s=2\pi/L$). The sum rule then reads
\be
\frac{\partial \hE}{\partial \beta}(n,\beta) = \square^{\Phi(n)}_\Omega \equiv \square^{n}_\Omega.
\la{eq:sr2}
\ee
We shall make use of the standard notation for discrete differences
\ba
\partial_n f(n) &=&  f(n+1) - f(n),\qquad \partial_n^* f(n) =  f(n) - f(n-1)\\
\widetilde\partial_n f(n) &=& \half(f(n+1)-f(n-1)) ,\quad \Delta_n f(n) = f(n+1) - 2f(n) + f(n-1).
\nonumber
\ea

Let us consider again the quantity $z = \hello \hE$,
which has a continuum limit. For definiteness take the case of a zero-momentum state
in finite volume, so that $L=na$. For a given couple $(n,\beta)$,
we choose an auxiliary $n'$ close to but different from $n$;
there is a $\beta'$ such that $n'/n= \hello(\beta')/\hello(\beta)$; that
is, the box sizes are matched in physical units. Then we can write
\ba
&& \hello \hE(n,\beta) = \hello(\beta') \hE(n',\beta') [1+{\rm O}(a^2)]  \\
&&= \hello \textstyle{\frac{n'}{n}} \left[\hE(n',\beta)+\dBdlogas \log\{n/n'\}
\square^{n'}_\Omega(\beta) + \half \log^2\{n/n'\} f(n',\beta) \right] [1+{\rm O}(a^2)]  
\nonumber
\ea
where $f(n',\beta)= \frac{d^2\beta}{d(\log a)^2} \db\hE(n',\beta) + \db^2E(n'\beta) (\dBdloga)^2 $. 
This second order term is necessary to eliminate ${\rm O}(a)$ cutoff effects.
We now choose $n'=n-1$ and then shift $n\to n+1$, obtaining Eq. (a);
secondly we choose $n'=n+1$ and then shift $n\to n-1$, obtaining Eq. (b).
Note that the function $f(n,\beta)$ is evaluated at the same arguments in (a) and (b).
Therefore we can eliminate that term by taking the linear combination 
$\log^2(1-1/n)\cdot(a)-\log^2(1+1/n)\cdot(b)$. This results in:
\be
\dBdlogas~ \square^n_\Omega [1+{\rm O}(a^2)]  = \hE(n) ~+~ n\widetilde\partial_n \hE(n) 
\la{eq:result3}
\ee
where we dropped the $\beta$ dependence and used the freedom to trade
one discretization scheme for another that is equivalent to ${\rm O}(a^2)$.
This is one of the Michael-Rothe sum rules~\cite{michael2,rothe1}. 
We have however kept track of the discretization errors carefully 
so as not to introduce ${\rm O}(a)$ effects. This sum rule allows one 
to extract the derivative of the torelon energy with respect to its length
without having to perform an independent simulation. In particular the effective
central charge can be computed via
\be
c_{\rm eff}(L=na,\hello) =  \half\textstyle{\dBdloga} \square^n_\Omega-n\hE(n) .
\la{eq:ceff1}
\ee
\subsection{The static potential case}
The sum rule~(\ref{eq:sr2}) also holds if static charges are inserted at points ${\bf 0}$
and ${\bf x}$, ${\bf x} = r\hat k$. Indeed the kernel of the transfer matrix then projects 
onto states $\Phi_{\alpha\beta\dots}$ which, under a gauge transformation 
$U\to U^\Lambda$, transform with $\Lambda({\bf 0})$ and $\Lambda({\bf x})$ 
according to the representation of these charges. 
For a fundamental-antifundamental pair, we have explicitly:
\be
\Phi_{\alpha\beta}[U^\Lambda] = (\Lambda({\bf 0})_{\alpha\gamma})^*  ~ 
 \Lambda({\bf x})_{\beta\delta} ~ \Phi_{\gamma\delta}[U].
\ee
The transfer matrix kernel is
\be
K_{\bf N^*\otimes N}^{\alpha\beta\gamma\delta}
= \int \prod_{\bx x} \dd W(\bx)~ \exp\left(-\beta\Delta S[V,W,V']\right)
  (W({\bf 0})_{\alpha\beta})^* ~ W({\bf x})_{\gamma\delta}
\ee
and the composition rule is 
\be
({\bf T}^2_{\bf N^*\otimes N})^{\alpha\beta\gamma\delta} = 
({\bf T}_{\bf N^*\otimes N})^{\alpha\lambda\gamma\epsilon} ~ 
({\bf T}_{\bf N^*\otimes N})^{\lambda\beta\epsilon\delta}.
\ee
Thus the partition function in the presence of the static charges is 
\be
{\cal Z}_{\bf N^*\otimes N} =\frac{1}{N^2} \sum_{\alpha,\gamma}
\tr\{ ({\bf T}_{\bf N^*\otimes N}^{\breve L_0})^{\alpha\alpha\gamma\gamma} \}
\ee
In particular, the Polyakov loop correlator evaluates to~\cite{lw-bosonic}
$\<P({\bf 0})^* ~ P({\bf x}) \> = {\cal Z}_{\bf N^*\otimes N} / {\cal Z}$.

The energies at 
two different separations $r$ must be subtracted in order to remove the 
divergent self-energy of the static quarks. 
The quantity $\hello^2(\beta) \partial_n \hE(n,\beta)$
has a continuum limit. We define $\bar n = n + \half + {\rm O}(a^2)$
where the ${\rm O}(a^2)$ need not be specified presently,
and choose two integers $n, n'\gg1,~|n-n'|={\rm O}(1)$ and a particular value of $\beta$.
Now there exists an auxiliary $\beta'$ such that
\[  \overline {n} a(\beta) = \overline {n'} a(\beta'), 
  \quad {\rm i.e.}\quad \overline {n} \hello(\beta') = \overline{ n'} \hello(\beta). \]
We can write $ \hello^2(\beta) \partial_n \hE(n,\beta) =
\hello^2(\beta') \partial_{n'} \hE(n',\beta') [1+ {\rm O}(a^2)]$ and then 
Taylor-expand in $\beta$ to obtain in the same way as in the previous section
\be
n\Delta_n\hE(n) + 2\widetilde\partial_n\hE(n) = 
\dBdloga \square^{n+1}_{n-1}~[1+{\rm O}(a^2)].
\ee
We use the notation $\square^{n+1}_{n-1}\equiv \square^{n+1}_\Omega - \square^{n}_\Omega$.
Again a symmetric finite-difference scheme is necessary and sufficient 
to remove the ${\rm O}(a)$ discretization errors.
\subsubsection{An improved estimate of $r_0^2F(r)$}
We now show how one can also use the lattice sum rule
to reduce cutoff effects on the static force. Discretization errors
are by far the dominant source of uncertainty on this quantity~\cite{lw-bosonic}.
\be
\ello\doteq r_0 \qquad {\rm and} \qquad z(\xi,\hro) \doteq \hro^2 F(r=\xi r_0,\hro).
\ee

A direct measurement of the force is 
\be
F(\overline n a, \hro) = \partial_n E(n,\hro).
\ee
This define the static force at a discrete set of points. 
To define it for all distances, 
an interpolation formula between neighbouring points must be used. 
Which formula is a matter of choice (different definitions 
of the force at finite lattice spacing will then differ by ${\rm O}(a^2)$).

For illustration we choose a linear interpolation between the two nearest 
direct measurements of the static force:
\be
z(\xi,\hro) = \frac{\hro^2}{\overline n - \overline {n-1}} 
\bigg[   (\xi\hro-\overline{n-1})  \partial_n \hE(n,\hro)
       + (\overline n - \xi\hro) \partial_n^* \hE(n,\hro)
\bigg]
\ee
With this precise definition and  \eq\ref{eq:sr2}, 
$\partial z/\partial\hro$ can be evaluated at fixed $\xi$ exactly.
An improved estimate of the continuum $z$ reads
\be
z_{\rm impr}(\xi,\hro) = z(\xi,\hro) + \half \hro \frac{\partial z}{\partial\hro},
\ee
which is improved in the same sense as the glueball mass in section~\ref{sec:impr}.
Explicitly, using \eq\ref{eq:sr2}, we get
\ba
z_{\rm impr}(\xi,\hro) &=& 2z(\xi,\hro) + \frac{\half\hro^2}{\overline n - \overline {n-1}} ~ \times\\
&& \bigg[ \xi \hro  \Delta_n E(n,\hro)  
-\dBdlogas\Big(  (\xi \hro-\overline{n-1})\square^{n+1}_n
               + (\overline n - \xi \hro) \square^{n}_{n-1}    \Big) \bigg]  \nonumber
\la{eq:zimpr2}
\ea
It is important to realize that the improvement is not merely a higher order difference
scheme for the static force, rather it contains non-perturbative information
about its lattice spacing dependence. Formula (\ref{eq:zimpr2}) may be numerically useful 
since all terms on the right-hand side can be evaluated in the same simulation.
Note that the idea is applicable to more complicated interpolation schemes
for any particular definition of $\overline n$, and also to 
the effective central charge $c_{\rm eff}(r)=-\half r^3d^2V/dr^2$.

\subsection{Anisotropic couplings\la{eq:aniso}}
Sum rules in the pure gauge theory derived by varying the lattice spacings in the four 
space-time directions independently around the isotropic point
 can be found in~\cite{michael2}.
We have little to add to this subject, except to say that 
the sum rules involving an external scale, such as a lattice size dependence,
must be expressed with a symmetric finite-difference scheme to avoid ${\rm O}(a)$ 
discretization errors. Here is a nice application: for a torelon in the direction $\hat1$,
the energy density in the transverse plane is directly sensitive to 
quantum string corrections, as already noted in~\cite{michael2},
and an effective central charge can be defined via
\be
c_{\rm eff}(L=na,\hello) = \frac{n}{2}~(U-S)~ 
\Big[ \square_{02}+\square_{03}-2\square_{23}\Big|^n_\Omega .
\la{eq:ceff2}
\ee
The definition of the anisotropic derivatives $U$ and $S$ 
will be given in section~\ref{sec:anisof}; we note that 
$(U-S)/2=\beta-N(c_\sigma-c_\tau)$, where the coefficients $c_\sigma,~c_\tau$ 
were determined in~\cite{karsch} (see also Ref. therein) for a few $\beta$ values.
The effective central charge extracted in this way will differ by ${\rm O}(a^2)$
terms from \eq\ref{eq:ceff1}.
\section{Lattice sum rules with Wilson fermions}
We start by recalling some essential facts about Wilson fermions
and their transfer matrix.
We shall use $\bx,\by$ for three-component spatial vectors,
indices $\alpha,\beta,\dots$ for color and $\rho,\sigma,\dots$ for spinor indices.
A spinor component is written as $\psi_\sigma(\bx, \alpha)$; for simplicity we 
consider the one-flavour theory for the moment, but the extension to several flavours
is trivial, as we shall see.
We use a set of Euclidean Dirac matrices, $\{\gamma_\mu,\gamma_\nu\}=2\delta_{\mu\nu}$, 
as well as the projectors $P_\pm = \half (1\pm\gamma_0)$.
Whenever there is a risk of confusion, we use a hat to distinguish 
a quantum mechanical operator from a Euclidean c-number or Grassmann variable.
The full Hilbert space ${\cal H}$ is now the tensor product of the gauge Hilbert 
space ${\cal H}_G$ and a fermionic Hilbert space ${\cal H}_F$.

An important object is the color-covariant, 
nearest-neighbour transport operator for fields 
transforming in the fundamental representation:
\be
D^\pm_k(\bx,\alpha;\by,\beta) =  U_k(\bx)_{\alpha\beta}~ \delta_{\bx + \hat k,\by}
               ~\pm~ U_k^\dagger(\by)_{\alpha\beta} ~\delta_{\by + \hat k,\bx},\qquad
k = 1,2,3.
\ee
Two spatial finite-difference operators will be used:
\ba
B &=& {\bf 1} - \kappa \sum_k D^+_k   \\
C &=&  \half \sum_k D_k^- \otimes \gamma_k.
\ea
$B$ is hermitian and strictly positive for $0<\kappa<1/6$,
while $C$ is anti-hermitian , 
for any gauge field configuration~\cite{luscher-transfmat}.

The fermionic Hilbert space is the Fock space built from a collection
of $\hchi$ operators, which enjoy canonical anticommutation relations
($\{\hchi_\sigma(\bx, \alpha),~\hchi_{\sigma'}^\dagger(\by, \beta)\} =
\delta_{\bx\by} ~ \delta_{\sigma\sigma'} ~ \delta_{\alpha\beta}$ etc.).
One further defines~\cite{luscher-transfmat}
\be
\hpsi_\sigma      =  B^{-1/2} ~\hchi_\sigma \qquad\qquad
\hpsi_\sigma^\dagger =  \hchi_\sigma^\dagger ~(B^{-1/2})^t,
\la{eq:chipsi}
\ee
that turn out to be the operators associated with the 
Grassmann variables $(\bar\psi,\psi)$.

The transfer matrix in the temporal gauge $A_0=0$ was obtained 
in terms of the $\hchi$ operators in~\cite{luscher-transfmat,creutz}.
The kernel of the transfer matrix in the path integral form 
is given in~\cite{sint-SF}, Eq. 4.26:
\ba
K[V,\bar\psi,\psi,V',\bar\psi',\psi'] &=& 
\det\{BB'\}^{1/2} \int \prod_{{\bf x}}
\dd W({\bf x}) ~ e^{-\Delta S[V,W,V']} \\
&&\exp\{-2(\bar\psi P_+ \kappa C\psi-\bar\psi P_+\kappa W^{-1}\psi'
 - \bar\psi' WP_-\kappa \psi + \bar\psi' P_{-} \kappa C'\psi')\},
\nonumber
\ea
where $B',C'$ are functionals of $V'$. The scalar product of two
functionals of $(\bar\psi,\psi)$ is taken with respect to 
Grassman integration with a measure $e^{-\bar\psi B\psi}/\det B $.
The determinant arises because of the change of basis (\ref{eq:chipsi}).

We can now derive a sum rule in the same fashion as in the pure gauge case. 
Let $\Phi$ be a normalized eigenstate of the transfer matrix with eigenvalue $\lambda$.
We find\footnote{It may be useful to note that if $T(x,y)=T^*(y,x)$,
$\lambda=\int w(x)dx w(y)dy \phi^*(x) T(x,y) \phi(y)$ and $\int \phi^*(x)\phi(x) w(x) dx = 1$, 
where $T$, $\phi$ and $w$ all depend on a parameter $\kappa$, then 
$\dk\lambda=\int w(x)dx w(y)dy \phi^*(x) (\dk T(x,y)) \phi(y) + \lambda \int (\dk w(x))dx |\phi(x)|^2$.}
\ba
\dk \lambda &=& \dk \< \Phi |  {\bf T}(\beta,\kappa) |\Phi \> =
2\int DV D\bar\psi D\psi~ \det\{B\}^{-1/2} ~ e^{-\bar\psi B\psi}~\Phi[V,\bar\psi,\psi]^*  \nn
&& 
\int DV' D\bar\psi' D\psi' ~ \det\{B'\}^{-1/2}  ~ e^{-\bar\psi' B'\psi'}~\Phi[V',\bar\psi',\psi']  \nn
&&\int DW ~ e^{-\Delta S[V,W,V']}~~ \Big( \bar\psi P_{+} W^{-1}\psi' + \bar\psi' WP_{-} \psi \Big)~\times\nn
&&\exp\{-2(\bar\psi P_+ \kappa C\psi - \bar\psi P_{+}\kappa W^{-1}\psi'
 - \bar\psi' WP_-\kappa \psi + \bar\psi' P_{-} \kappa C'\psi')\}  \nn
&+&\lambda \int DV D\bar\psi D\psi~ \frac{e^{-\bar\psi B\psi}}{\det B}  |\Phi[V,\bar\psi,\psi]|^2
\Big( \bar\psi(\textstyle{\frac{1}{\kappa}}(1-B)-2C)\psi   \Big)
\la{eq:k-deriv}
\ea
In the last line, $\lambda$ multiplies the expectation value on the state $\Phi$ of an 
equal-time operator $\hat A_s$, 
while the rest of the expression is the expectation value on the state $\Phi$
of an integral operator $\hat A_t$.
Following the discussion of the pure gauge case, we thus have
\be
-\dk \hE(\beta,\kappa) = \< \Phi | \hat A_s|\Phi \> -\< \Omega | \hat A_s|\Omega \>
          + e^{\hE}   \< \Phi | \hat A_t|\Phi \> - e^{\hE_{\Omega}}  \< \Omega | \hat A_t|\Omega \>
\equiv {\rm h}|^\Phi_\Omega.
\la{eq:result-f}
\ee
To evaluate this matrix element in a Monte-Carlo simulation, 
one would use an interpolating field $\varphi$ for the state $\Phi$. If it is 
the lightest in its symmetry channel,
\be
\< \Phi | \hat A_{s}|\Phi \> = \lim_{x_0\to\infty} f_{s}(x_0), \qquad\qquad
e^{\hE} \< \Phi | \hat A_{t}|\Phi \> = \lim_{x_0\to\infty} f_{t}(x_0) \qquad
\ee
with (for $x_0/a$ even)
\be
\textstyle{
f_s(x_0)=\frac{\<\bar\varphi(0)\frac{1}{2}(A_s(x_0/2)+A_s(x_0/2+a)) 
\bar\varphi(x_0+a)\>}{\<\bar\varphi(0) \bar\varphi(x_0+a)\>},
\quad
f_t(x_0)=\frac{\<\bar\varphi(0)A_t(x_0/2) \bar\varphi(x_0+a)\>}{\<\bar\varphi(0) \bar\varphi(x_0+a)\>}
}
\ee
and
\ba
A_s(x_0) &=& \sum_{{\bf x},k} \bar\psi(x) U_k(x)(1-\gamma_k) \psi(x+a\hat k)
                   +\bar\psi(x+a\hat k) U_k(x)^{-1}(1+\gamma_k) \psi(x)      
\la{eq:A_s} \\
A_t(x_0) &=& \sum_{\bf x}\bar\psi(x) U_0(x) (1-\gamma_0) \psi(x+a\hat0)
                        + \bar\psi(x+a\hat0) (1+\gamma_0) U_0(x)^{-1}\psi(x).
\la{eq:A_t}
\ea
These are the hopping terms of the Wilson-Dirac action; 
they are the terms one would obtain by naively differentiating the Boltzmann factor. 
It is also clear now that equations (\ref{eq:k-deriv}) and (\ref{eq:A_s}, \ref{eq:A_t}) hold
for $N_{\rm f}$ degenerate flavours, provided an implicit 
summation over flavours is understood in (\ref{eq:A_s}, \ref{eq:A_t}).
In practice, once the fermion fields are integrated out, 
the three-point functions involve all-to-all propagators
within one time-slice. A stochastic estimator is then required, as used 
in previous thermodynamics applications~\cite{CP-PACS}.

\subsection{Applications}
We now present some applications of the sum rule just established.
We focus on the case of $N_{\rm f}$ degenerate flavours.
Let us assume for now the spatial volume to be large enough for the state $\Phi$
to suffer negligible finite volume effects (for a stable one-particle state,
they are exponentially small). Given the existence of the continuum limit,
we have, up to cutoff effects,
\be
0 = \frac{d}{da} [\frac{\hE }{a} (\beta(a),\kappa(a))] \qquad \Rightarrow \qquad 
\hE = \partial_\beta \hE \dBdloga + \partial_\kappa \hE \dkdloga.
\ee
But actually, $\kappa= \kappa(\beta(a))$: on a line of `constant physics'
$\kappa$ must be tuned as a function of $\beta$. Hence
\be
\hE = \dBdloga \left[ \square^\Phi_\Omega  -~ \dkdb ~{\rm h}|^\Phi_\Omega \right][1+{\rm O}(a)].
\la{eq:af}
\ee
where we have used \eq\ref{eq:result} and \eq\ref{eq:result-f}.
Recall that $\dBdloga \sim -4Nb_0$ is universal in the continuum limit
($b_0=\frac{1}{3(4\pi)^2}(11N-2N_{\rm f})$).
A particularly interesting case 
arises for $N_{\rm f}\geq2$ in the chiral limit, $\kappa=\kappa_c$.
The `pion' mass then vanishes and hence 
\be
\frac{d\kappa_c}{d\beta} = \frac{\square^{\pi}_\Omega}{{\rm h}|^{\pi}_\Omega}
\cdot [1+{\rm O}(a)]
\la{eq:kc}
\ee
Note that at no stage in the derivation of the sum rules did we need to make 
an assumption on the extent of the time direction -- only the spatial lattice size 
was assumed to be essentially infinite. Hence this equation also holds in the 
Schr\"odinger functional -- where one can actually simulate at $\kappa_c$.
Relation (\ref{eq:kc}) can now be inserted into the sum rule for the only other stable 
particle, the nucleon:
\be
\hM_{\rm nucl} = \dBdloga\Big[\square^{\rm nucl}_\Omega - 
\frac{{\rm h}|^{\rm nucl}_\Omega}{{\rm h}|^{\pi}_\Omega}
\square^{\pi}_\Omega \Big]~[1+{\rm O}(a)].
\la{eq:pn}
\ee

As in the pure gauge theory, we can give an 
improved estimate of the continuum $z\equiv \hello\hM$, 
assuming now that the leading corrections
to the spectrum are ${\rm O}(a)$. Here $\hello(\beta,\kappa_c(\beta))$ is the quantity 
evaluated in the chiral limit; concrete examples are $r_0|_{m_\pi=0}$, 
$\Lambda_{\overline{\rm MS}}^{-1}$ or $L_{\rm max}$, defined by $\bar g^2(1/L_{\rm max})=x$,
where $x$ is a particular numerical value and $\bar g^2$ 
the renormalized Schr\"odinger functional coupling~\cite{nf2alpha}. 
We get
\be
z_{\rm impr}(\hello) \equiv   z(\hello) + dz/d\log\hello
= 2z(\hello) -
\hello\dBdloga \Big(\square^\Phi_\Omega - \dkdb~{\rm h}|^\Phi_\Omega \Big)
\ee
We expect the ${\rm O}(a)$ cutoff effects to be substantially reduced on this estimator. 
Note that for this formula to yield an improvement, $\dBdloga $ must be defined 
through the quantity $\ello$, as in~\eq(\ref{eq:dBdloga}).

If the volume dependence of the energy level cannot be neglected, 
\eq\ref{eq:af} becomes
\be
 \hE(\{\breve L_k\}) + \sum_k \breve L_k \widetilde\partial_{\breve L_k} \hE= \dBdloga
\Big[ \square -~ \dkdb ~{\rm h} \Big|^{\Phi(\{L_k\})}_\Omega~ [1+{\rm O}(a)].
\la{eq:afv}
\ee
The derivation is identical to the one for the corresponding relation in the 
pure gauge case.
The finite-difference scheme used here in the derivative is formally irrelevant, 
since the sum rule holds with ${\rm O}(a)$ corrections anyhow; however we still
expect the symmetric scheme to yield somewhat smaller discretization errors.
This formula may be useful to study the volume dependence of the pion mass, 
and also of the state which is a mixture of a two-pion state and 
the $\rho$ resonance. The latter volume dependence 
allows one to extract the scattering lengths of pions in the elastic regime
and the $\rho$ width~\cite{luscher91}.

\subsection{Anisotropic lattice   \la{sec:anisof}}
We consider the Wilson theory with $N_{\rm f}$ degenerate flavours.
If one assigns an independent hopping parameter
in the space directions from the time direction,
then one can derive the equations
\ba
-\frac{\partial \hE}{\partial\kappa_\tau} &=& e^{\hE} \<\Phi|\hat A_t|\Phi\>
                          - e^{\hE_\Omega} \<\Omega|\hat A_t|\Omega\> \equiv {\rm h}_t|^\Phi_\Omega \\
-\frac{\partial \hE}{\partial\kappa_\sigma} &=&  \<\Phi|\hat A_s|\Phi\>
                          -  \<\Omega|\hat A_s|\Omega\> \equiv {\rm h}_s|^\Phi_\Omega
\ea

To  exploit these relations in a useful way,  we must 
understand how the theory with parameters $(\beta_s,\beta_t,\kappa_s,\kappa_t)$
is renormalized. As independent `physical' variables, the most intuitive choice is
$(\hello,\xi,x)$, where $\xi$ is the anisotropy $a_s/a_t$ and $x$ essentially sets
the quark masses (e.g. $x=m_{\rm PS}/m_{\rm V}$). The continuum limit is then 
taken with $\hello^{-1}\to0$ at fixed $\xi$ and $x$. One can alternatively choose the
set $(a_t,a_s,x)$ as independent variables. 
Since there are four lattice parameters, it is only on a hypersurface
in that parameter space that the lattice theory describes QCD at all. 
The continuum limit corresponds
to a particular curve on that hypersurface where $\xi$ and $x$ are constant.

Imagine momentarily using four independent lattice spacings,
so that $(a_0,a_1,a_2,a_3,x)$ are a set of independent variables.
Introduce one hopping parameter $\kappa_\mu$ for each direction,
$\beta_{\mu\nu}$ for each plaquette orientation 
and let $G\equiv \partial\kappa_0/\partial \log a_0$ 
and $H\equiv\partial\kappa_0/\partial\log a_k$. 
Now locking {\small $a_1=a_2=a_3=a_s$},
the following relations hold at the isotropic point {\small $a_s=a_t$}:
\ba
\partial\kappa_k/\partial a_t = H &\qquad\qquad&
\partial\kappa_k/\partial a_s = G+2H \nn
\partial\kappa_0/\partial a_t = G &\qquad\qquad&
\partial\kappa_0/\partial a_s = 3H.
\ea
Similarly, if $S=\partial \beta_{0k}/\partial\log a_0 $ 
and $U\equiv \partial \beta_{kl}/\partial\log a_0$, then 
in the {\small $a_1=a_2=a_3=a_s$} theory, we have at the isotropic point 
{\small $a_s=a_t$}~\cite{michael2}:
\ba
\beta_{0k}/\partial\log a_t = S &\qquad\qquad&
\beta_{0k}/\partial\log a_s = S+2U  \nn
\beta_{kl}/\partial\log a_t = U  &\qquad\qquad&
\beta_{kl}/\partial\log a_s = 2S+U
\ea
In this way the number of functions has been reduced by a factor two.
However, to our knowledge the derivatives $U,S,G,H$ have not yet been 
determined individually in the $N_{\rm f}=2$ theory.

By writing the renormalization group equations
$\frac{d}{da_{s,t}} \frac{\hE}{a_t}=0$, and taking suitable linear combinations
thereof, one recovers \eq\ref{eq:af} (or more generally \eq\ref{eq:afv}) 
under the consistency conditions
\be 
2(S+U)=\dBdloga \qquad\qquad G+3H = \frac{d\kappa}{d\beta},
\ee
and obtains the independent sum rule
\be
\hE-{\textstyle\frac{1}{3}}\sum_k \breve L_k\widetilde\partial_{\breve L_k}\hE 
= -\bigg\{ {\textstyle\frac{2}{3}}(U-S) \Big[\sum_k \square_{0k} 
- \sum_{k<l} \square_{kl}\Big|^{\Phi}_\Omega
+{\textstyle\frac{1}{3}}(G-H) \Big[3{\rm h}_t-{\rm h}_s\Big|^\Phi_\Omega\bigg\}
[1+{\rm O}(a)].
\la{eq:anisof}
\ee
The vacuum matrix elements vanish in this case by space-time symmetry.

In thermodynamic studies of 
QCD using the Wilson regularization~\cite{CP-PACS}, 
the energy density $\epsilon$ 
and the pressure $p$ are extracted in the so-called derivative 
method by computing derivatives with respect to $\beta$ and $\kappa$ 
of the partition function. Since ${\cal Z}=\tr\{{\bf T}^{\breve L_0}\}$, where
the temperature $T$ is equal to $L_0^{-1}$, one finds that 
$aV(\epsilon-3p)$, the `interaction measure', is given by 
a thermal average (with Boltzmann measure $e^{-E/T}$)
over the eigenstates of the transfer matrix 
of expression (\ref{eq:afv}). Similarly the expression corresponding to 
$aV(\epsilon+p)$ is \eq(\ref{eq:anisof}).

\section{Conclusion}
We have extended the known set of lattice sum rules by including
the effects of Wilson fermions. Our derivation in the transfer matrix
formalism kept track of discretization errors. The main results are 
\eq(\ref{eq:afv}) and (\ref{eq:anisof}).

We have also presented new applications of the sum rules.
Since they allow us to compute the derivative of 
an energy level with respect to the lattice spacing,
we have proposed to exploit this to reduce cutoff effects on the spectrum
and on the static potential. One may  include the information on the slope
in the continuum extrapolation, if one disposes of data at several lattice 
spacings. Alternatively, one can form an improved estimate of the continuum limit
from data at a single lattice spacing
under the assumption of a functional form for the leading discretization
errors, such as the one predicted by Symanzik's effective theory~\cite{symanzik}.

Of course these applications are only of practical interest if the three-point
functions can be computed accurately in numerical simulations. 
This can probably be achieved for the static potential 
in the pure gauge theory using a suitable multi-level algorithm~\cite{inprep}. 
It also seems realistic in
$N_{\rm f}\geq2$ theories in the pseudoscalar sector, although 
high statistics and efficient all-to-all techniques will be needed.
We remark that higher order sum rules may be derived straightforwardly 
by taking additional derivatives of the transfer matrix with respect 
to the bare parameters, but the $n$-point functions are bound to become 
more difficult to evaluate numerically with increasing $n$.
It is nevertheless conceptually pleasing that locally the lattice artefacts
can be determined at fixed bare lattice parameters.

\paragraph{}
I am happy to thank Karl Jansen and Rainer Sommer for reading and commenting
on the manuscript.
{\small

}
\end{document}